\documentclass{jfm}
\usepackage[dvips]{graphicx}
\usepackage{amsmath}
\usepackage{amssymb}
\usepackage{setspace}
\usepackage{psfrag}
\usepackage{color}
\usepackage{graphicx}
\usepackage{natbib}
\usepackage{multirow}
\usepackage{mathrsfs}

\definecolor{AJWpurple}{rgb}{0.3,0.3,0.3}
\definecolor{AJWdarkgreen}{rgb}{0.5,0.5,0.5}
\definecolor{AJWHead}{rgb}{0.2,0.2,0.7}

\title[Finite sample size effects on convection in mushy layers] {Finite-sample-size effects on convection in mushy layers}
\author[J. -Q. Zhong, A. Fragoso, A. J. Wells  and J. S. Wettlaufer]
{J.-Q.\ns  Z\ls H\ls O\ls N\ls G$^{1}$,\ns A.\ns T.\ns F\ls R\ls A\ls G\ls O\ls S\ls O$^{1,2}$,\ns A.\ns J.\ns W\ls E\ls L\ls L\ls S$^{1,3}$\ns and J.\ns  S.\ns W\ls E\ls T\ls T\ls L\ls A\ls U\ls F\ls E\ls R$^{1,2,3,4}$}

\affiliation{$^{1}$Department of Geology and Geophysics, Yale University, New Haven, CT, USA \\
$^{2}$Department of Physics, Yale University, New Haven, CT, USA\\
$^{3}$Program in Applied Mathematics, Yale University, New Haven, CT, USA\\
$^{4}$NORDITA, Roslagstullsbacken 23, SE-10691 Stockholm, Sweden}

\begin{document}

\maketitle 

\begin{abstract}
We report theoretical and experimental investigations of the flow instability responsible for the mushy-layer mode of convection and the formation of chimneys, drainage channels devoid of solid, during steady-state solidification of aqueous ammonium chloride. Under certain growth conditions a state of steady mushy-layer growth with no flow is unstable to the onset of convection, resulting in the formation of chimneys.
We present regime diagrams to quantify the state of the flow as a function of the initial liquid concentration, the porous-medium Rayleigh number, and the sample width. For a given liquid concentration, increasing both the porous-medium Rayleigh number and the sample width caused the system to change from a stable state of no flow to a different state with the formation of chimneys. Decreasing the concentration ratio destabilized the system and promoted the formation of chimneys. As the initial liquid concentration increased, onset of convection and formation of chimneys occurred at larger values of the porous-medium Rayleigh number, but the critical cell widths for chimney formation are far less sensitive to the liquid concentration. At the highest liquid concentration, the mushy-layer mode of convection did not occur  in the experiment. The formation of multiple chimneys and the morphological transitions between these states are discussed.  The experimental results are interpreted in terms of a previous theoretical analysis of finite amplitude convection with chimneys, with a single value of the mushy-layer permeability consistent with the liquid concentrations considered in this study.   
\end{abstract}

\section{Introduction\label{sec1:intro}}

The solidification of binary alloys and the formation of so-called mushy layers are important in applications across geophysics, geology and industry, ranging from sea ice growth~\citep{Feltham:2006gf}, to solidification in the Earth's interior~\citep{Bergman:1994kx}, to the formation of metal castings~\citep{Copley:1970qt}. Mushy layers are partially-solidified reactive porous materials comprised of solid crystals and interstitial melt, and formed as a result of morphological instability during the solidification of a two component melt~\citep{MullinsSekerka:1964}. An issue of significance in many applications is understanding the conditions for both the occurrence and the resulting dynamics of buoyancy-driven convection within the mushy region~\cite[see][for a review.]{Worster:1997} Convective flow in a mushy layer transports fluid into regions of differing concentration, which can lead to local solidification or dissolution of the solid matrix and changes in the effective material properties. The permeability of the porous matrix depends on the local solid fraction, and hence flow focussing can accelerate the growth of the underlying convective-flow instability, and eventually lead to the formation of chimneys -- channels of zero solid fraction that provide efficient pathways for buoyancy-driven flow out of the mushy layer. Understanding the onset and dynamics of convection in mushy layers with chimneys is of importance in geophysics for predicting brine fluxes from sea ice~\citep{Wettlaufer:1997rt,Wellsetal:2011}, and in metallurgy because material heterogeneities associated with chimneys  can lead to defects in metal castings~\citep{Copley:1970qt}.

Experimental studies of chimney formation in mushy layer growth typically take one of two forms. Numerous so-called fixed-chill experiments have studied the transient growth of a mushy layer from a cold isothermal boundary and show that chimneys typically form after the mushy layer exceeds a critical thickness~\cite[see, for example,][and references therein]{Aussillousetal:2006}. As the thickness of the mushy layer continues to increase, it is observed that the mean spacing between chimneys also increases, with extinction of the flow in certain chimneys~\citep{Wettlaufer:1997rz,Solomon:1998}. Alternatively one also considers steady-state mushy layer growth. Directional solidification experiments, where the sample material is translated between two heat exchangers, indicate that multiple chimneys form for growth rates slower than a critical threshold that depends on the initial liquid concentration~\citep{Peppinetal:2008,Whiteoak:2008kx}. \cite{NeufeldWettlaufer:2008a} used growth from a boundary with controlled time-dependent temperature in order to achieve steady-state growth and study the influence of an external shear flow on the onset of chimney formation. For both transient and steady state growth the critical threshold for chimney convection can be interpreted in terms of an appropriate mushy-layer Rayleigh number exceeding a critical value, so that buoyancy effects overcome dissipation and lead to the onset of convective fluid flow. 

Theoretical modelling of convective flow in mushy layers has been approached from a variety of angles. Linear and weakly nonlinear stability analyses have considered the onset of convection during directional solidification under a wide variety of conditions~\cite[see, for example,][and references therein for a review and more recent summaries.]{Worster:1997,GUBA:2010rz,Roperetal:2011} Several studies have revealed that the bifurcation to both steady and oscillatory convective flow states can be subcritical~\citep{AmbergHomsey:1993,Anderson:1995zl,GUBA:2006zl} depending on the growth conditions. Hence, this raises the possibility that finite amplitude perturbations may allow states with chimney convection to be accessed even when the base state of no flow is linearly stable. Convective states of finite amplitude have been considered either in direct simulations of particular growth conditions~\cite[][]{Felicellietal:1998,Beckermannetal:2000,Guo:2003fv,Heinrich:bh,Oertling:2004rz,Jain:2007rt,Katz:2008tg} or via studies of the system dynamics with an assumed periodic array of chimneys~\citep{SchulzeWorster:1998,ChungWorster:2002,Wellsetal:2010}. In a study of particular relevance here \cite{Wellsetal:2010} used a numerical approach to consider the influence of changes in the spacing between chimneys on the evolution of system dynamics and the stability of solutions with steady-state chimney convection. This approach demonstrates that the observed convective flow states are suppressed via a saddle-node bifurcation as the chimney spacing is reduced.

In this study we focus on the influence of a finite sample width on the formation of chimneys during mushy layer convection. In \S\ref{sec2: theoretical modelling} we describe how the theoretical model of directional solidification of \cite{Wellsetal:2010} can be used to study the dynamics and stability of nonlinear convective flow states of finite amplitude. 
An experimental procedure is described in \S\ref{sec3:apparatus} that produces steady-state mushy layer growth for comparison to the theoretical predictions of \citet{Wellsetal:2010}. An extensive set of experiments described in~\S\ref{sec4:results} allows us to identify conditions for chimney formation for a variety of sample widths, growth rates and liquid concentrations, with comparison to the theoretical results providing an indirect estimate of mushy layer permeability for several different liquid concentrations. We conclude with a discussion in \S\ref{sec6:discussions}.

\section{A model of mushy layer convection with chimneys\label{sec2: theoretical modelling}}

A theoretical model of mushy layer convection with chimneys was used by \cite{Wellsetal:2010} to investigate the dynamics of nonlinear convective flow, and is applied here to determine conditions under which flow states with a chimney can exist. We briefly summarise the essential features of the model below, noting that further details of the governing equations and boundary conditions are elaborated on in \cite{Wellsetal:2010}. 

\subsection{Physical and theoretical framework \label{sec:theory}}

\begin{figure}
\centering
\psfrag{LIQUID}{\textsc{Liquid}}
\psfrag{MUSH}{\textsc{Mush}}
\psfrag{SOLID}{\textsc{Solid}}
\psfrag{CHIMNEY}{\textsc{Chimney}}
\psfrag{L}{$l$}
\psfrag{V}{$V$}
\psfrag{azt}{$a(z,t)$}
\psfrag{z}{$z$}
\psfrag{x}{$x$}
\psfrag{0}{$0$}
\psfrag{zeqhxt}{$z=h(x,t)$}
\psfrag{TCPhiPsi}{$T$, $C$, $\phi$, $\hat{\mathbf{u}}$}
\psfrag{CeqCETeqTE}{$C=C_E,\quad T=T_E$}
\psfrag{CeqCoTeqTo}{$C=C_o,\quad T=T_{\infty}$}
\includegraphics[height=6.0cm]{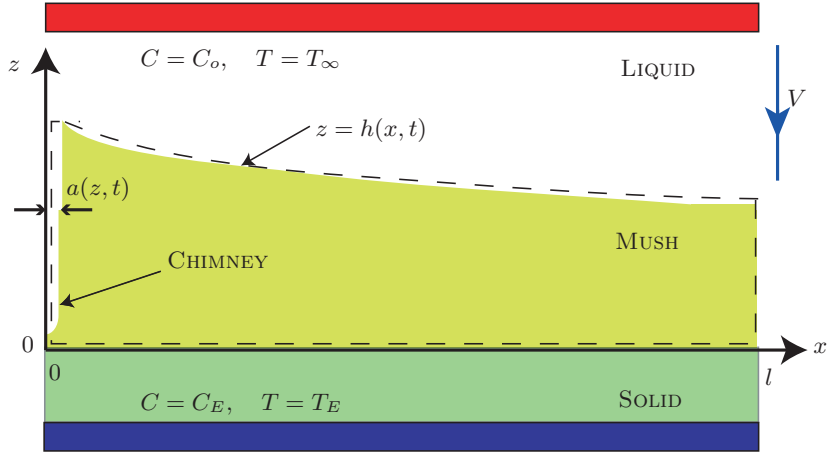}
\caption{Schematic of the theoretical framework.  Liquid of initial concentration $C_o$ and temperature $T_{\infty}$ is translated downward at speed $V$ between hot and cold heat exchangers, and partially solidifies to form a mushy layer of thickness $h(x,t)$. We adopt a local co-ordinate system $(x,z)$ so that a eutectic solid of temperature $T_E$ and concentration $C_E$ forms at $z=0$. We solve in a domain $0\leq x\leq l$ and $0\leq z\leq h(x,t)$ as indicated by the dashed outline, with a chimney of width $a(z,t)$ located at $x=0$ located in this domain. With suitable boundary conditions,  described in the main text, this domain describes either a periodic array of convection cells separated by chimneys of width $2a(z,t)$ centred at $x=0$ and with inter-chimney spacing $2l$, or else a single convection cell with chimney of width $a(z,t)$ against an impermeable free-slip vertical boundary.}
\label{fig:notation}
\end{figure}

The model considers the directional solidification in two dimensions of a binary solution of local temperature $T$ and local concentration $C$, translated  at a speed $V$  between hot and cold heat baths, as illustrated in figure~\ref{fig:notation}.
 This results in a porous mushy region of thickness $h(x,t)$ bounded by an impermeable eutectic solid with temperature $T_E$ and and concentration $C_E$ at $z=0$, and an overlying fluid for $z>h$ with assumed uniform composition $C_o$ and far-field temperature $T_{\infty}$. We simulate flow with a chimney and single convection cell of width $l$, utilizing boundary conditions that simultaneously allow us to consider both convection cells that are either isolated, or in periodic arrays. The temperature $T$, local solid fraction $\phi$ and Darcy velocity $\mathbf{u}$ within the mushy region are calculated using time-dependent ideal mushy layer theory~\citep{SchulzeWorster:2005}  which conserves heat and salt, accompanied by a condition of local thermodynamic equilibrium which couples the temperature and liquid concentration via a freezing point relationship $T=T_L(C)$. We use a linear approximation $T_L(C)\equiv T_E+\Gamma(C-C_E)$ for the freezing point relationship,  where $\Gamma$ is constant. Fluid velocities are determined from the incompressible Darcy flow law, where the permeability is assumed to be $\Pi=\Pi_0 (1-\phi)^3$ for some reference permeability $\Pi_0$. We apply symmetry conditions on $T$ and no normal flow at $x=l$, equivalent to either an insulating, impermeable boundary or a periodic array of convection cells. 

The pure liquid flow in the narrow chimney is driven by buoyancy forces or pressure gradients and is parameterised using lubrication theory~\cite[see][and references therein]{Wellsetal:2010} where we apply symmetry conditions on the velocity $\mathbf{u}$ at $x=0$. This velocity boundary condition can describe either flow within two convection cells separated by a chimney of width $2a(z,t)$ centred at $x=0$, or else a single convection cell with chimney of width $a(z,t)$ against an impermeable free-slip vertical boundary (a discussion of the likely influence of using a free-slip condition rather than a no-slip condition follows at the end of~\S\ref{sec:theory}). 
Combining the lubrication approximation with similar asymptotic approximations for heat transport yields nonlinear boundary conditions on the left hand side of the mushy region, which is treated as a singular interface at $x=0$. These conditions depend on the width of the chimney, with $a(z,t)$ updated to satisfy a condition of marginal equilibrium to just remove any constitutional supercooling. The boundary conditions at the mush liquid interface and procedure for updating $h(x,t)$ are described in more detail in~\cite{Wellsetal:2010}. A pertinent point is that this approach uses a boundary layer approximation that describes the circulation and heat transport in the overlying fluid that is induced by the mushy-layer mode of convection, but neglects any component of buoyancy-driven convection that arises due to density differences within the overlying fluid region.

This model is characterized by six dimensionless parameters,
\begin{align}\lambda&=\frac{V l}{\kappa},  &  \mathrm{R_m}&=\frac{\rho_0 g\beta(C_0-C_E)\Pi_0}{\mu V}, &  \mathscr{C}&=\frac{C_S-C_0}{C_0-C_E},  \nonumber \\
 \theta_{\infty}&=\frac{T_{\infty}-T_L(C_{0})}{\Gamma (C_0-C_E)},& \mathcal{S}&=\frac{L}{c_p\Gamma (C_0-C_E)}, & \mathrm{Da} &= \frac{\Pi_0 V^2}{\kappa^2}, \label{eq:params2} \end{align}
 where the fluid density is approximated by $\rho_0\left[1+\beta(C-C_E)\right]$ for constant haline coefficient $\beta$ and reference density $\rho_0$, $\mu$ is the dynamic viscosity, $g$ is the gravitational acceleration, $\kappa$ is the thermal diffusivity, $c_p$ is the specific heat capacity, $L$ is the latent heat of fusion, and we have made the simplifying assumption that material properties are uniform in both solid and liquid phases in line with ideal mushy layer theory~\citep{SchulzeWorster:2005}. The numerical results presented herein investigate how the dynamics change under variations of the dimensionless sample width $\lambda$ and mush Rayleigh number $\mathrm{R_m}$ which characterizes the strength of buoyancy effects compared to dissipation. The experimental concentration and temperature conditions are reflected in the concentration ratio $\mathscr{C}$, $\theta_{\infty}$,  and $\mathcal{S}$, with the Darcy number $\mathrm{Da}$ representing the relative permeability of the porous matrix. It is difficult to directly constrain and measure the permeability in experimental settings, and so in~\S\ref{sec:stabbound} we determine a reference value of $\Pi_0=1.6 \times 10^{-5} \,\mathrm{cm}^2 $ for a range of liquid concentrations, by comparison between experimental data and theoretical prediction. This provides an indirect estimate of the permeability for solidification in a  Hele-Shaw cell. Comparison to other estimates in different settings will be discussed in \S\ref{sec6:discussions}. Importantly, we note that an independent set of simulations that vary $\mathrm{Da}$ with all other parameters held fixed show that whilst the chimney width varies with scaling $aV/\kappa \sim \mathrm{Da}^{1/3}$ consistent with the scaling analysis of \cite{SchulzeWorster:1998}, the fluid flux remains constant at leading order over the range $10^{-4}\leq \mathrm{Da}\leq 5\times10^{-3}$. The same features have also been observed in a recently-developed simplified model of convection in mushy layers~(Rees-Jones \& Worster, personal communication). This behaviour is consistent with the flow dynamics being controlled by buoyancy-driven flow within the mushy region, whilst the chimney width adjusts passively to accommodate the necessary liquid flux. Hence, the mushy-region flow dynamics do not depend strongly on the ratio of flow resistance in the porous mushy region to viscous resistance in the liquid region, consistent with only a weak dependence on $\mathrm{Da}$. We maintain the value $\mathrm{Da}= 5\times10^{-3}$ throughout all numerical results reported here. 
 
The fact that flow in the mushy region is relatively insensitive to viscous resistance in the liquid region also suggests that qualitatively similar results would be obtained by application of a free-slip and a no-slip boundary condition in the chimney. The boundary condition describing flow at the chimney edge takes the form
\begin{equation}\psi=\alpha_1\frac{\mathrm{R}_m}{\mathrm{Da}} \left(\frac{aV}{\kappa}\right)^3\kappa\left[\frac{1}{3\mathrm{R}_m(1-\phi)^3}\frac{\partial \psi}{\partial x} + \frac{3\alpha_2}{20} V\frac{C-C_E}{C_0-C_E}\right]+ \alpha_3 a\frac{\partial \psi}{\partial x}, \label{eq:chimneypsi} \end{equation}
\citep{Wellsetal:2010}, where the streamfunction $\psi$ satisfies $\mathbf{u}=(-\psi_z,\psi_x)$  and the constant pre-factors are $\alpha_1=\alpha_2=\alpha_3=1$ for a free-slip boundary, or $\alpha_1=1/4$, $\alpha_2=2/3$ and $\alpha_3=1/2$ for a no-slip boundary. For fully developed chimneys, with $aV/\kappa\sim \mathrm{Da}^{1/3}$ the final term on the right-hand side of~\eqref{eq:chimneypsi} is $O(\mathrm{Da}^{1/3})$ smaller than the remaining terms~\citep{SchulzeWorster:1998}, and so an order-one change in $\alpha_3$ does not impact the leading order behaviour. Next, we note that the numerical coefficient of the buoyancy term (featuring $\alpha_2$) has been determined using a Polhausen approximation in the current calculations~\citep{Wellsetal:2010}. Previous calculations using a different form of approximation yield a different numerical coefficient, but qualitatively similar mushy-region-flow dynamics~\citep{ChungWorster:2002}. Hence, we expect an order-one change in $\alpha_2$ to also produce qualitatively similar dynamics. Finally, the difference in $\alpha_1$ between free-slip and no-slip boundaries can be compensated for by a rescaling of the Darcy number $\mathrm{Da}\rightarrow 4\mathrm{Da}$. Hence, because the flow in the mushy layer is insensitive to $\mathrm{Da}$ at leading order, we expect no qualitative difference in the leading-order dynamics within the mushy region between cases with a free-slip or a no-slip boundary condition in the chimney.

\subsection{Numerical method}

The solution procedure combines two different numerical approaches involving both time dependent calculations, and an arc-length-continuation scheme. The time dependent simulations combine second-order finite differences in space, with semi-implicit Crank-Nicholson timestepping, and free boundaries updated via relaxation~\cite[see][for further details]{Wellsetal:2010}. The resulting systems for temperature and streamfunction were solved using a multigrid iterative method~\citep{Adams:1989,BriggsHensonMcCormick:2000}. This time dependent code was complemented by an arc-length-continuation method~\citep{Keller:1977} that traces both stable and unstable solution branches, allowing us to verify that the relaxation procedures do not have a strong influence on the stability of the system. In order to reduce the computational costs in solving the systems involving Jacobian terms that are required for arc-length continuation, the predictor-corrector calculations are computed from projections of the finite difference solutions onto a basis of products of Chebyshev polynomials in $x$ and $y$. This allows the main characteristics of the solution to be characterised by a smaller number of degrees of freedom. The resulting solutions predicted by arc-length continuation in Chebyshev space are then refined using a steady version of the finite-difference code.

\subsection{Stability boundaries} 

\begin{figure}
\centering
\psfrag{Fs}{$F$}
\psfrag{lambda}{~~~~~~~~~~~~$\lambda$}
\psfrag{lsFs}{$(\lambda_c,F_c)$}
\includegraphics[height=9.55cm]{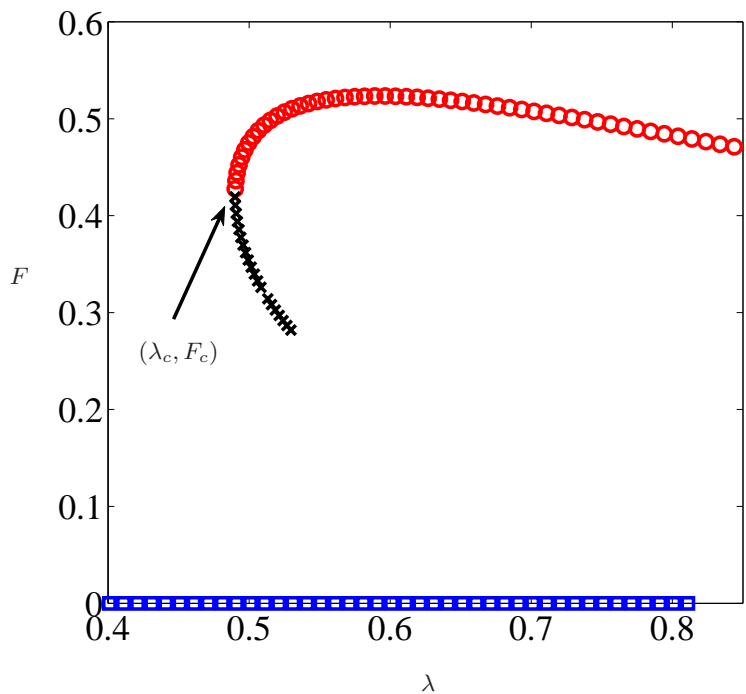}
\caption{Variation of the calculated dimensionless brine flux $F$ with convection cell width $\lambda$ for steady state flow with $\mathrm{R_m}=32.5$ $\mathscr{C}=11.7$, $\mathcal{S}=4.8$, $\theta_{\infty}=0.185$, and $\mathrm{Da}=0.005$, consistent with solidification of a $27 \,\mathrm{wt}\%$ aqueous ammonium chloride solution described in \S\ref{sec3:apparatus}. The system shows hysteresis, with multiple stable steady states. The stable upper branch of chimney convection (red circles) and stable lower branch of no flow (blue squares), are accompanied by an intermediate unstable branch of chimney convection (black crosses). The system undergoes a saddle-node bifurcation at $(\lambda_c,F_c)$, with the stable state of chimney convection ceasing to exist for $\lambda<\lambda_c$. Failure of convergence of the iterative corrector scheme prevents further continuation along the unstable branch.  }
\label{fig:solbranch}
\end{figure}

In this paper we are interested in the influence of sample width on the formation  of chimneys in a mushy layer. To determine conditions for the existence of states with chimneys, the arc-length-continuation method is used to trace solution branches as $\lambda$ varies with $\mathrm{R_m}$, $\mathscr{C}$, $\mathcal{S}$, $\theta_{\infty}$ and $\mathrm{Da}$ all held fixed, applying the method of~\cite{Wellsetal:2010} to a region of parameter space relevant to our experiments. Figure~\ref{fig:solbranch} shows a typical example of the system evolution, where $F$ is the dimensionless solute flux from the chimney. 
Two steady states are observed, an upper branch of chimney convection (red circles) and a lower branch of no flow (blue squares), along with an intermediate unstable branch of chimney convection (black crosses). Starting on the upper branch of stable chimney convection, as $\lambda$ is decreased the system undergoes a saddle-node bifurcation at $\lambda=\lambda_c$ so that the chimney convection state ceases to exist for $\lambda<\lambda_c$. Having determined the critical wavelength $\lambda_c$, the above procedure is repeated with varying $\mathrm{R_m}$ but  $\mathscr{C}$, $\mathcal{S}$, $\theta_{\infty}$ and $\mathrm{Da}$ held fixed to obtain a stability boundary of the form $\lambda=\lambda_c(\mathrm{R_m})$. Examples of such stability boundaries and their comparison to experimental data are shown later in figure~\ref{fig4} and discussed in~\S\ref{sec4:results}. Note that, by construction, the theory predicts that flow with a single chimney at the centre of two symmetric convection cells requires twice the domain width as the corresponding state with a single convection cell and chimney at the sidewall. Hence one intuitively expects that a state of convection with a single wall chimney can be achieved at smaller Rayleigh numbers and narrower cell widths than convection with chimneys away from the sample walls.

\section{Experimental apparatus and methods\label{sec3:apparatus}}
\begin{figure} 
\centering
\includegraphics[width=10cm]{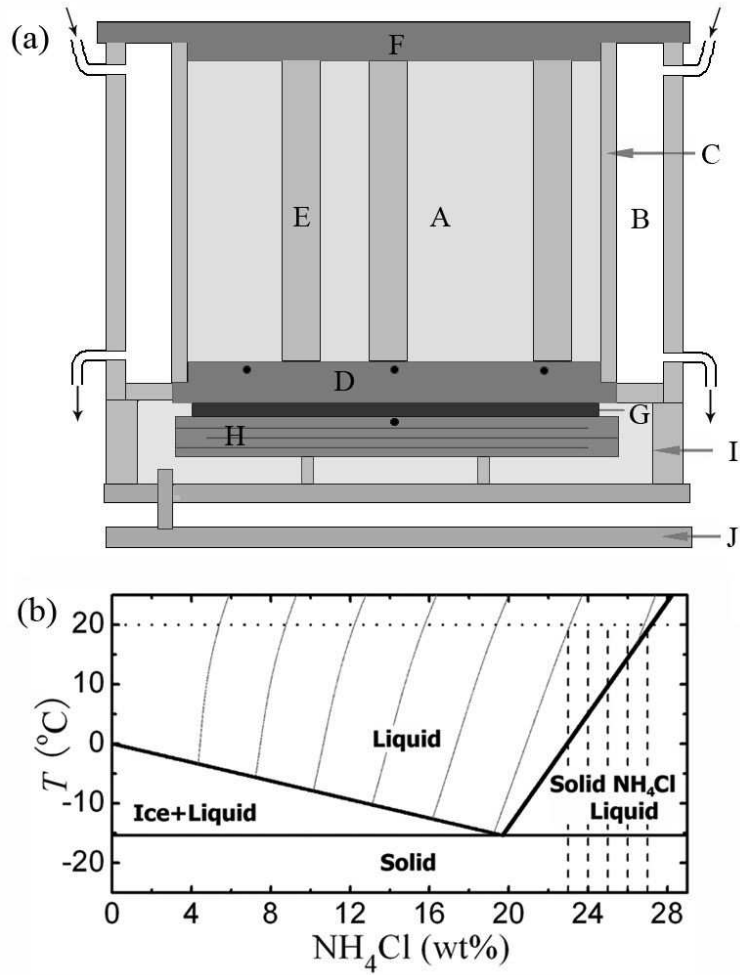}
\caption{(a) Schematic of the experiment apparatus. See text in Sec. 3.1 for detailed identification of the various components. (b) Phase diagram of the NH$_4$Cl-H$_2$O solution. The vertical dashed lines indicate the initial liquid concentrations chosen in the experiment. Solution density isopleths are indicated by the faint lines with positive slope above the liquidus and were constructed 
using data from \citep{InternationalCriticalTables:2003} }
\label{fig1}
\end{figure}

\subsection{Experimental apparatus\label{sec31:apparatus}}
A schematic diagram of the apparatus is shown in Fig. \ref{fig1}a. A Hele-Shaw sample container (A) of inner dimensions $25.40\times15.24\times0.32$ cm is surrounded by an air-flow vessel (B). The container walls (C) are made of 0.8-cm thick, optically clear plexiglas, and extend 0.6 cm below the top surface of the bottom plate (D). The seal which confines the working fluid is provided by high-strength silicon adhesive (RTV 732) below the walls. Spacers (E) made by low-thermal-conductivity plexiglas divide the container into a subset of cells with adjustable width $l$. The width of the spacers $l_s$=1.3 cm is chosen to provide sufficient thermal insulation between the convection cells. The thermal diffusion time over the spacers, $l_s^2/\kappa_s$ is of order hours ($\kappa_s$ is the thermal diffusivity of plexiglas),  and is comparable to the whole time span of the solidification process. In this manner, mushy-layers grow in each of the convection cells independently, and each is subjected to spatially uniform cooling from the bottom plate.

The cell bottom (D) is made from a 1.3-cm thick high thermal conductivity copper plate. Three thermistors are embedded in (D) with a 0.3-cm distance to the top plate surface (shown by the solid black circles in Fig. \ref{fig1}a). These monitor the bottom temperature of the sample $T_B$. To inhibit chemical reactions with the working fluid, the top surface of (D) is painted with an anti-corrosion coating (PRO 15) 0.1mm thick. When the highest thermal flux is applied through the sample, this coating is responsible for a temperature difference of 0.3K between $T_B$ and the fluid temperature measured at the bottom of the sample. Corrections have thus been made for $T_B$ to accurately reflect the bottom fluid temperature. 

The sample cell is enclosed by an air-flow vessel (B). Ambient air enters and leaves through outlets in this vessel, as 
indicated by the arrow heads in Fig. \ref{fig1}a, at a rate of approximately 50 cm$^3$/s. The air flow has been thermally stabilized at $T_{\infty}=20\pm0.2^{\circ}$C in a large reservoir before entering the vessel to provide a steady thermal environment. Simultaneously we flow dry air flow through this region to suppress water condensation on the sample walls (C) during the solidification process. An aluminium top plate (F) maintains a constant far-field temperature $T_{\infty}$ of the sample. The working fluid confined in the cell is an aqueous NH$_4$Cl solution with initial concentration $C_0$. The density and temperature of the solution are measured before the experiments using a oscillating-type densitometer (Antor Paar, DMA 35N). The concentration $C_0$ is then chosen between 23wt$\%$ and 27wt$\%$ with an accuracy better than $\pm$ 0.1wt$\%$ using the isopycnal lines as shown in the phase diagram in Fig. \ref{fig1}b.

In thin cells the thermal boundary conditions of the sample may be affected by heat flow through the walls to the working fluid, which has been considered explicitly in a recent mathematical model \citep{Peppinetal:2007}.  Because of the finite thermal resistance of the wall material, a sample cell with wall temperature profiles that match entirely the fluid temperature is necessary to eliminate the lateral heat flow effects. The double-sided wall construction in the present experimental system effectively minimizes this heat loss effect. For a quantitative evaluation of the lateral heat lost through the sidewalls, we measure the fluid temperatures by moving a temperature probe along the shortest dimension of the cell, at a horizontal level 3mm above the mush-liquid interface before the formation of the chimneys. The temperature variation is within 0.03K over the full thickness (0.32mm) of the cell, with the centre fluid temperature slightly higher than that near the sidewalls. Thus, we conclude that the vertical heat flux through the sample dominates the lateral heat lost along the sidewalls, and that the temperature field in the liquid region is essentially two-dimensional.  

Underneath the bottom of the sample cell there is an array of Peltier devices (G) (thermoelectric coolers) sandwiched between the copper plate (D) and a 2.5-cm thick copper manifold (H).  The temperature of the manifold is regulated through the circulation of anti-freezing coolant. High-conductivity thermal compound is pasted between the Peltier devices and the copper plates to enhance the thermal contacts and improve the thermal homogeneity on the bottom plate of the sample. Under typical experimental conditions, the horizontal temperature heterogeneity along the sample bottom is within 0.05K, as measured by the three thermistors embedded in part (D).

Supporting the copper manifold (H), a thermal insulating based shield (I) rests on a levelling plate (J) through which the entire apparatus can be levelled within 0.005 radian.

\subsection{Thermometry and thermal control protocol\label{sec32:thermal control}}

The bottom-plate temperature $T_B$ and top-plate temperature $T_{\infty}$ are measured with thermistors imbedded in the bottom plate (D) and the top plate (F). The fluid temperature in the liquid region is measured by thermistors inserted inside the sample through 0.6-mm diameter stainless steel tubing. All thermistors are calibrated to a precision of 1mK against a Hart Scientific model 5626 platinum resistance thermometer and are traceable to the ITS-90 temperature scale. Consistency between the thermistors is within 0.01K.

Recent experiments \citep{Peppinetal:2007, Peppinetal:2008} investigated steady state mushy-layer growth using an experimental apparatus in which the solution is translated at prescribed rates between two heat exchangers. In the present experiment, steady growth is accomplished by time dependent temperature control as $T_B=T_B^{\circ}-{\gamma}t $ where ${\gamma}$ is the constant cooling rate.

\begin{figure}
\centering
\includegraphics[width=10cm]{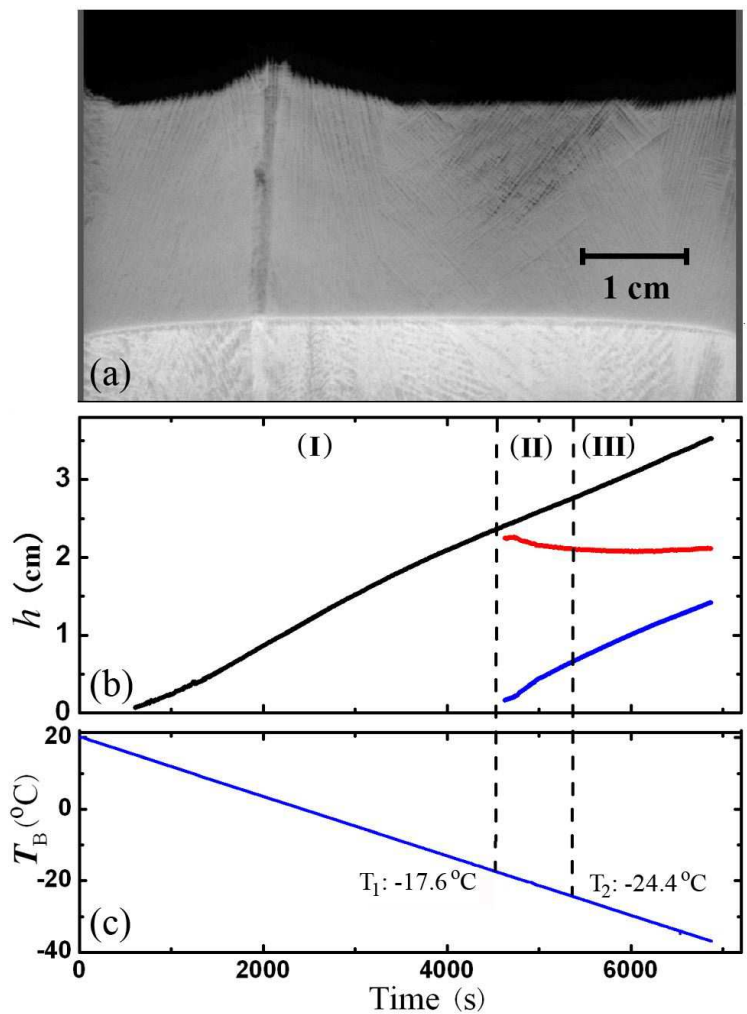}
\caption{(a) Image of a mushy-layer growing from a 27wt$\%$ NH$_4$Cl solution with cooling rate $\gamma$=0.5K/min. The bright white curve is the eutectic-mush interface. The horizontal black bar is 1 cm. (b) Time-recording of the mushy-layer thickness depicting three growth periods. Black curve: height $H_I$ of the mushy-liquid interface; Blue curve: height $H_E$ of the eutectic solid layer; Red curve: thickness of the mushy-layer $h=H_I-H_E$. (I) Linear-growth period. In this period the bottom-plate temperature $T_B$ is above the eutectic temperature $T_E$. The mushy-layer grows linearly as a function of time until the eutectic solid forms. (II) The initial transient-growth period ($T_1>T_B>T_2$). The eutectic layer starts to grow in this period. The mushy-layer thickness  depends on time and finally converges to a constant. (III) Steady-growth period ($T_B<T_2$). The mushy-layer thickness remains constant. (c) The linearly decreasing bottom-plate temperature $T_B$. The two dashed lines in (b) and (c) show the two temperatures at state transitions: $T_1=-17.6^{\circ}$C,  $T_2=-24.4^{\circ}$C. }
\label{fig2}
\end{figure}

Following the protocol developed in a recent experimental study, control of the sample bottom-plate temperature $T_B$ is achieved by two cooling mechanisms  \citep{NeufeldWettlaufer:2008a}. First, we use an array of  Peltier devices (part (G) in Fig. \ref{fig1}a) to produce a thermal flux from the bottom plate (D). The downward heat flux, which is proportional to the applied electrical power to the Peltier devices, is controlled in a feedback loop that incorporates the instantaneous temperature readings for $T_B$, and is delivered to the Peltier devices by a programmable power supply (Sorenson, Model DCS 40-25). A second ingredient of the cooling-control settings is the coolant-circulating copper manifold (H). Anti-freeze coolant is circulated in a double-spiral channel built inside manifold (H), driven by two Neslab (RTE-7) baths at a flow rate of 500 cm$^3$/s and a cooling capacity of 1000 Watt at 20$^{\circ}$C. The bath-temperature stability achieved over long time scales is within 0.01$^{\circ}$C, as measured by the thermistor embedded in the manifold (H). The Peltier devices provide fast control of the temperature variation on the bottom plate (D). During the experiment a prescribed linearly decreasing bottom-plate temperature $T_B$ of the sample is maintained down to approximately -40$^{\circ}$C (Fig. \ref{fig2}c), which forces steady growth of the overlying mushy layer as depicted in Fig. \ref{fig2}b.

\section{Results\label{sec4:results}}

\subsection{Steady state mushy-layer growth}

\begin{figure}
\centering
\includegraphics[width=10cm]{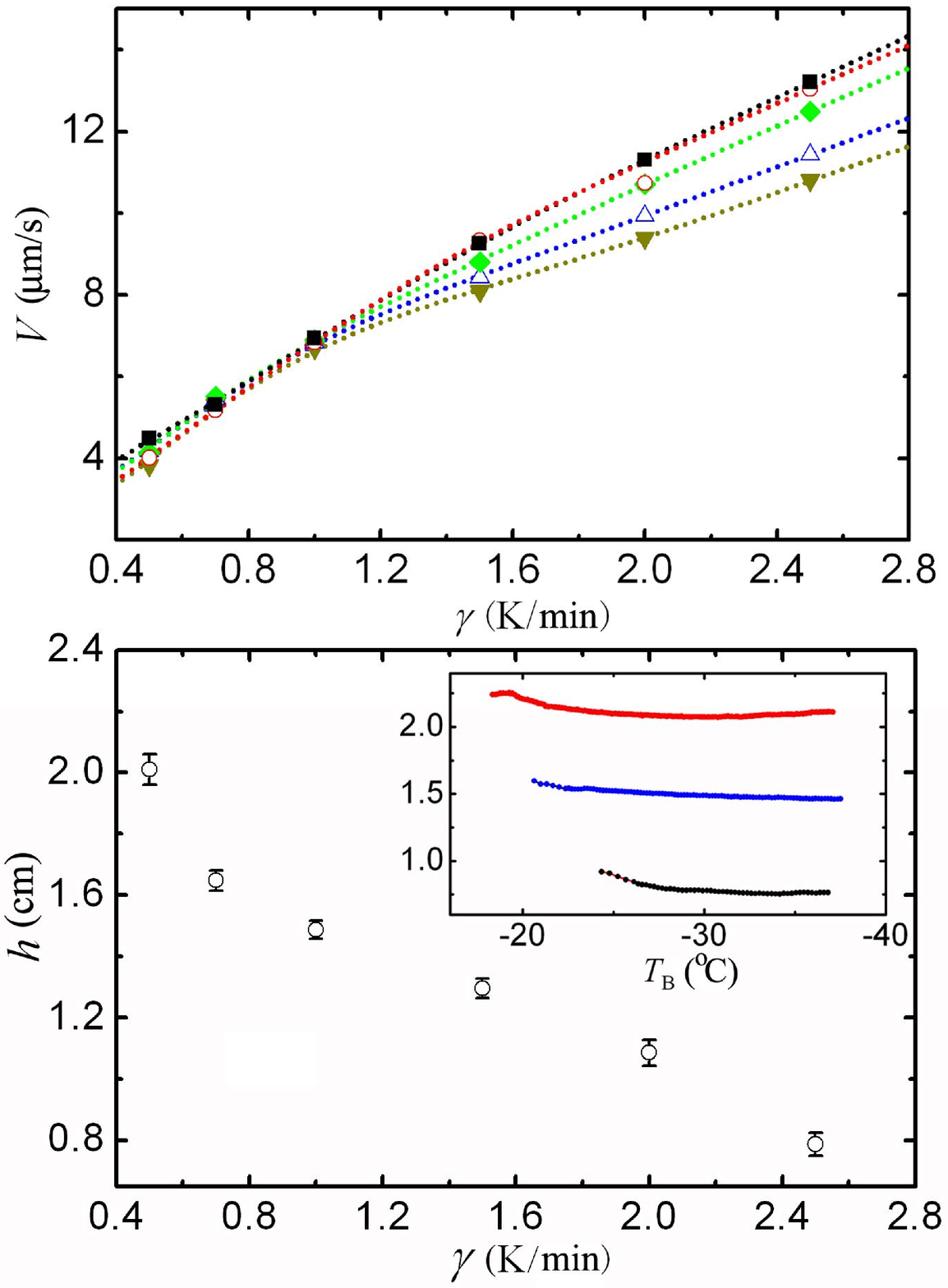}
\caption{(a) Growth velocity $V$ of the mush-liquid interface as a function of the  cooling rate $\gamma$ in a steady-growth state with $l=2.54$ cm. Black squares: $C_0=27\,\text{wt}\%$;  Red circles: $C_0=26\,\text{wt}\%$;  Green diamonds $C_0=25\,\text{wt}\%$;  Blue  triangles: $C_0=24\,\text{wt}\%$; Dark yellow triangles: $C_0$=23wt$\%$. Dashed lines: spline fitting curves for each set of data points. (b) Height of the mushy-layer $h$ as a function of the cooling rate $\gamma$ in a steady-growth state with $C_0=27\,\text{wt}\%$, $l=2.54$cm. The error bars show the variation in $h$ at different runs under the same experimental conditions. Inset: The mushy-layer height $h$ as a function of $T_B$ at three cooling rates. Red curve: $\gamma$=0.5K/min; Black curve: $\gamma$=1.0K/min; Blue curve: $\gamma$=2.5K/min. } 
\label{fig3}
\end{figure}
One representative set of data for the mushy-layer growth as a function of time is shown in Fig. \ref{fig2}b. There are three distinct periods, with different characteristic variations of $h(t)$ with time. To the left of the first dashed line at $T_B=T_1=-17.6^{\circ}$C (shown in Fig. \ref{fig2}c), the thickness of the mushy layer increases linearly with a growth speed of $ 6.60 {\mu}$m/s. No eutectic layer is formed in this period. Formation of a eutectic solid layer underlying the mushy-layer starts at $T_B=T_1$. As the bottom-plate temperature $T_B$ further decreases an initial transient-growth period appears between ($-17.6^{\circ}$C, $-24.4^{\circ}$C), as indicated by the two vertical dashed lines. Growth of the mushy-layer in this period is dependent on initial conditions, particularly the cooling history before $T_B$ reaches $T_1$. 

To the right of the second dashed line at $T_B=T_1=-24.4^{\circ}$C, both the advancing speed of the mush-liquid interface, $\dot{H}_I$, and the eutectic-mush interface, $\dot{H}_E$, are constant and equal $V= 4.86{\mu}$m/s. The thickness of the mushy-layer, $h=H_I-H_E$, thus remains a constant independent of the initial cooling process. The thickness of both the mushy layer and the eutectic layer in this steady-growth period are shown in Fig. \ref{fig2}a. 

From the description above it is clear that the advancing speed of the mush-liquid interface $V$ is not a prescribed control parameter in this study but depends on the bottom-plate cooling rate $\gamma$ and the initial liquid concentration $C_0$. In Fig. \ref{fig3}a we show that at high cooling rates, when $\gamma >1.0$K/min, $V$ increases with $C_0$, whereas at low cooling rates $\gamma\le1.0$K/min, $V$ is nearly independent of $C_0$.     

In Fig. \ref{fig3}b we show data for the mushy-layer thickness $h$ during a steady state of chimney convection at cooling rates covering $0.5$K/min $\le \gamma \le2.5$K/min. The dependence of $h$ as a function of the bottom-plate temperature is shown in the inset for three cooling rates $\gamma=$0.5, 1.0 and 2.5K/min. When $T_B$ is sufficiently low the system easily reaches a steady state. The mushy-layer thickness $h$ is then evaluated when it converges to a constant and is independent of  $T_B$.  Figure \ref{fig3}b shows that the mushy-layer tends to be thicker at lower $\gamma$, leading to a higher mushy-layer Rayleigh number. These trends agree qualitatively with theoretical predictions \citep{Worster:1992b}.

Results for $h$ at the same cooling rates $\gamma$ are repeatable from run to run and are independent of the pre-cooling process. The inter-experimental variation of the steady-state mushy-layer thickness $h$  is within $5\%$, as indicated by the error bars in Fig. \ref{fig3}b.  Experiments with very low cooling rate $\gamma<$0.5K/min show that the advancing speed of the eutectic-mush interface is always larger than the mushy-layer interface ($\dot{H}_E > \dot{H}_I$), leading to a decreasing mushy-layer thickness and the system does not reach a steady state within the experimental timeframe.  Due to the finite cooling capacity of the experimental apparatus, at experiments with high cooling rate $\gamma>$2.5K/min, the bottom-plate temperature $T_B$ deviates from the usual linearly decreasing trends in the low-temperature range.  Data analysis in these experiments is limited by the shorter period of  growth.

\subsection{Instability and onset of mushy-layer mode convection \label{sec:stabbound}}

\begin{figure}
\centering
\includegraphics[width=9cm]{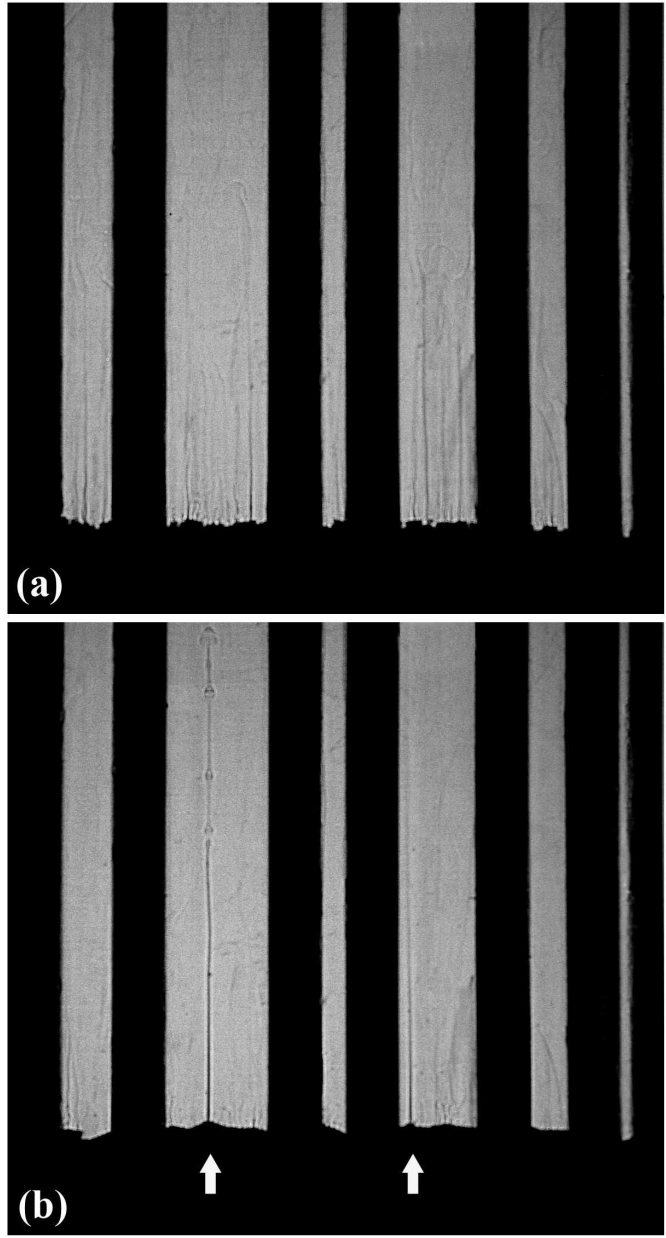}
\caption{Shadowgraph images captured for $C_0=25\,\text{wt}\%$ and (a)  $\gamma$=1.5K/min;  (b) $\gamma$=1.0K/min. Stability analysis results based on such images are presented in the regime diagrams in Fig. 7b  by downward and upward triangles, respectively. Both pictures are taken at $T_B=-40.0^0$C. Note the formation of a strong chimney plume in each of the two largest cells in (b), indicated by two white arrows. The horizontal white bars show 2 cm.} 
\label{fig5}
\end{figure}

\begin{figure} 
\centering
\includegraphics[width=13cm]{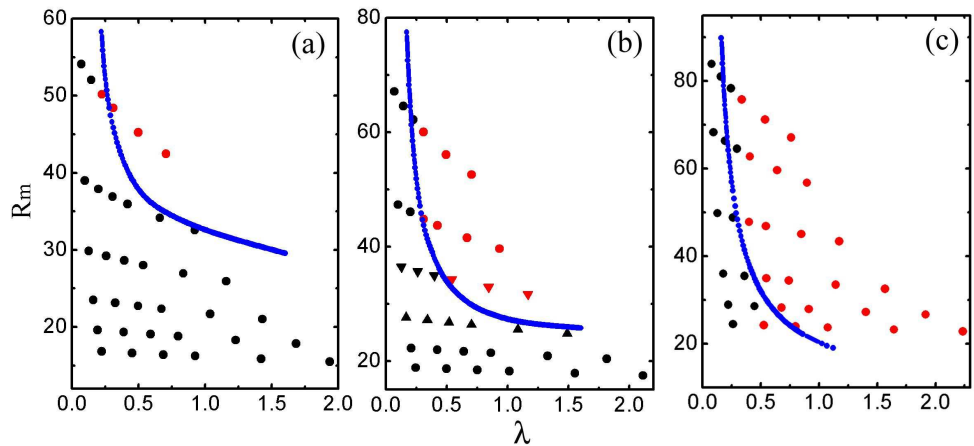}
\caption{Regime diagrams depicting the stability of the mushy-layer mode convection. Black symbols: mushy-layer grows in a stable state. Red symbols: unstable state with formation of at least one chimney channel within the mushy-layer. (a) Results for  $C_0=24\,\text{wt}\%$. (b) Results for $C_0=25\,\text{wt}\%$. The triangles indicate the experimental results illustrated by shadowgraph pictures in Fig. \ref{fig5}. (c) Results for  $C_0=27\,\text{wt}\%$. In addition, similar experiments for $C_0=23\,\text{wt}\%$ (not shown) resulted in no chimney formation for this range of cell widths and Rayleigh numbers. Experimental values of the Rayleigh number are calculated using the reference permeability ${\Pi}_0=1.6{\times}10^{-5}$ cm$^2$ for all three liquid concentrations, determined as described in the main text. Blue curves: numerical results for the stability curve indicating the critical R$_m^c$ as a function of $\lambda$. }
\label{fig4}
\end{figure}

For mushy-layer Rayleigh numbers just above a critical value R$_m^c$, we observed the formation of chimneys, dendrite-free regions within the mushy-layer (Fig. \ref{fig2}a). The instability controlling the transition from the no-chimney state to the mushy-layer mode of convection depends on the mushy-layer Rayleigh number,   dimensionless cell width $\lambda=Vl/\kappa$ in addition to $\mathscr{C}$, $\mathcal{S}$ and $\theta_{\infty}$ which depend on the liquid concentration $C_0$.  To identify a convective state in which chimneys form, we capture shadowgraph images to illustrate two-dimensional fluid density variations for visualization of compositional plumes emerging from the mushy layer. We require a state of chimney convection to satisfy three criteria. First, there is a channel of zero solid fraction within the mushy region (Fig. \ref{fig2}a). Second, there is at least one vigorous compositional plume in the far-field liquid region. Since intermittent chimney plumes \cite[or convection in a breathing mode,][]{Solomon:1998, Peppinetal:2008} are commonly observed in the experiment, we consider a chimney plume if its life time is longer than the diffusion time scale for boundary layer convection  $d^2/D\sim10$ mins. Here $d\sim1$mm is the average size of an isolated plume detached from the compositional boundary layer in the liquid region and $D$ is the solute diffusivity. Third, other turbulent buoyant plumes in the near field next to the chimney plume have become substantially diminished which indicates suppressed boundary-layer mode convection in concert with the formation of the chimney. Transition from a state of no flow to a state of chimney convection is readily identified through shadowgraph pictures. In Fig. \ref{fig5}, we see the flow patterns in the liquid region when two sets of  parameters ($\lambda$,R$_m$) are chosen. Figure \ref{fig5}a shows that in a stable state of no flow (these data points are indicated by upward triangles in Fig. \ref{fig4}b), there is no chimney flow in the far field of the liquid region in any of the cells, while small-scale turbulent plumes driven by the boundary-layer mode convection remain active. When the value of R$_m$ increases slightly  (indicated by downward triangles in Fig. \ref{fig4}b), chimney flow penetrates into the far field in the two largest cells (Fig. \ref{fig5}b), which are readily discernible from the small-scale turbulent plumes and dominate the flow field in the liquid region. We thus conclude that the mushy-layers growing in the two largest cells are both in an unstable configuration that leads to a state of chimney convection. 

We determine the permeability of the mushy layer  $\Pi_0$ in the following manner. By adjusting a single value of $\Pi_0$ that is the same for three different liquid concentrations, we determine values of $\mathrm{R}_m$ from the experimental data so that the boundary between the stable and unstable regimes is closest to the stability curve derived from numerical calculations. As illustrated in Fig. \ref{fig4}, the resulting comparison with the theoretical curves is consistent with the observed stability to within one increment in cell width $\lambda$. Hence, this comparison between theory and experiment allows us to empirically estimate the permeability as $\Pi_0=1.6{\times}10^{-5}$ cm$^2$, independent of liquid concentration in the range  $24\,\text{wt}\%\leq C_0\leq27\,\text{wt}\%$. 

As seen in Fig. \ref{fig4}a, mushy-layer growth at a low initial liquid concentration ($C_0=24\,\text{wt}\%$) is mostly stable. Chimneys are present only at high values of  R$_m$ and large $\lambda$. Fig \ref{fig4}b shows that as the liquid concentration increases (to $C_0=25\,\text{wt}\%$), the critical R$_m^c$ decreases.  Moreover, the sensitivity of R$_m^c$ to $\lambda$ increases, abruptly increasing as the confinement increases (i.e. as $\lambda$ decreases) and preventing the onset of mushy-layer mode convection.  When the liquid concentration further increases  to $C_0=27\,\text{wt}\%$, solidification of the sample with large $\lambda$ becomes unstable for all the values of R$_m$  obtained from the experiments as shown in Fig. \ref{fig4}c.  In summary, the system becomes less stable as the liquid concentration $C_0$ increases, which can be interpreted as follows. An increase in liquid concentration $C_0$ leads to the values of the concentration ratio $\mathscr{C}$, the Stefan number $\mathcal{S}$, and the dimensionless liquid temperature $\theta_{\infty}$ all being reduced.  Changes in these parameters influence the stability of the system in different ways, as identified in the linear stability analysis of~\cite{Worster:1992b}. If $\mathscr{C}$ is reduced, then the mean solid fraction is higher and hence the local permeability is reduced within the interior of the mush.  In addition, a reduced value of $\mathscr{C}$ leads  to a thinner mushy layer and reduces the potential energy that is available to drive flow. Both of these effects make the system more stable. A smaller value of $\mathcal{S}$ implies that the influence of latent heat is smaller, so that the solid matrix is dissolved more easily. This promotes growth of the instability. A smaller value of $\theta_{\infty}$ leads to a smaller heat flux from the liquid, so that the mushy layer grows thicker and provides greater potential energy to drive flow. The present theoretical simulations of finite amplitude convection reveal that the net combination of these competing effects results in the system becoming less stable as the liquid concentration $C_0$ is increased.

\begin{figure}
\centering
\includegraphics[width=13cm]{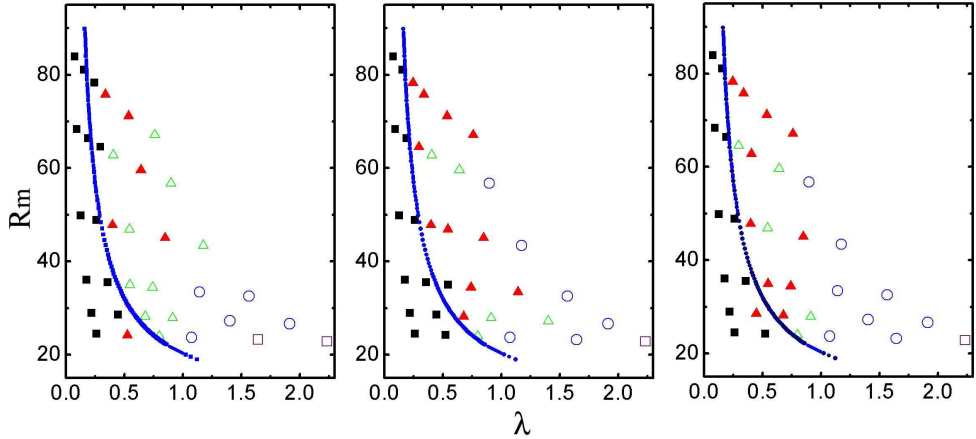}
\caption{Results for the morphological states of the mushy-layer mode convection. Data obtained from three sets of experiments with $C_0=27\,\text{wt}\%$. Black solid squares: stable state, mushy-layer with no-chimney; Red solid triangles: one chimney forms at the walls; Green open triangles: one chimney forms off the walls; Blue open circles: formation of two chimneys within the mushy-layer; Purple open squares:  formation of three or more chimneys within the mushy-layer. Blue curves: numerical results for the stability curve indicating the critical R$_m^c$ as a function of $\lambda$ (corresponding to a single convection cell with chimney at the wall.)}
\label{fig6}
\end{figure}

\subsection{Variability, multiple chimneys and morphological transitions between states}

Depending on the growth speed of the mushy-layer and the size of the sample-cell, several morphological transitions are observed in the states displaying chimney convection.  To illustrate these morphological transitions  we present
 these results  as regime diagrams for three repetitions of experiments with $C_0=27\,\text{wt}\%$ in Fig. \ref{fig6}. The comparison between the three repetitions in panels (a), (b) and (c) is indicative of the role of fluctuations in influencing different experimental realisations. The stability boundary between states with and without flow is robust to within inter-experimental variation of one increment amongst the set of cell widths. The variability between observed states with and without flow under the same experimental conditions might be explained in part by the theoretically suggested sub-critical instability behaviour: the form of perturbation required to trigger a transition to a finite-amplitude convecting state might only be achieved in a subset of the experiments. The variability may also be influenced by factors not described in the theory, such as weak heat loss through the walls or boundary-layer-mode convection. 

The regime diagrams in Fig. \ref{fig6} also reveal a preference for chimney formation at the cell walls when the system undergoes a transition from a state of no flow to a state of chimney convection at $\lambda\equiv\lambda_{c,1}$. Combining all the data in figure \ref{fig6}, for each cooling rate we determine the states at the smallest sample width that permits convection. Using this method we find four cell widths where a single chimney is observed at the cell wall, and the remaining two cell widths show either a chimney at the wall, or a chimney in the sample interior, for different repetitions under the same conditions. Localization of the mushy-layer mode of convection near the walls has been observed in both experiments and models \citep{Giameietal:1970, Felicellietal:1998}. Recent theoretical calculations show that chimney formation can be influenced by lateral heat transfer due to heat losses at the sidewalls. Sidewall heat transfer can have both direct thermal effects, and indirect effects by forming weak convective flows in the overlying fluid \citep{Roperetal:2007, Roperetal:2011}. The cooling effect of sidewall heat losses can promote the formation of chimneys near to the walls~\citep{Roperetal:2007}. Whilst localization of convection resulting from flow in the overlying fluid encourages chimney formation at the walls in three-dimensional mushy-layer cells, in two-dimensional flow the localisation promotes chimney formation away from the cell walls. Hence, this form of flow localisation cannot be the controlling factor in our quasi-two-dimensional Hele-Shaw cell. As an alternative factor, our theoretical model suggests that the preference for wall chimneys close to the critical cell width for onset of convection may simply be a geometrical effect due to the finite-sample-width effects in our experiment. If the stability boundary $\lambda=\lambda_{c,1}$ for states with one chimney at the wall is interpreted as the minimum width of a single convection cell, then the narrowest sample widths that can support convection necessarily have a chimney at the wall. A state with a central chimney  separating two symmetric convection cells requires a minimum cell width width $\lambda=2\lambda_{c,1}$. Whilst we have not conducted detailed stability calculations for asymmetric states where an interior chimney is located away from the centre of the sample, it is plausible that certain forms of these states might be stable over the range $\lambda_{c,1}<\lambda<2\lambda_{c,1}$ with the chimney separating a dominant convection cell of width greater than $\lambda_{c,1}$ from a narrower passive region. The possibility of asymmetric arrangements of convection cells is also likely to be of relevance in describing the observed transitions to states with two or more chimneys.  For the same value of R$_m$, transition from one-chimney states to two-chimney states occurred at a large spacing $\lambda_{c,2}$. Although the regime boundary for this transition varies slightly between different runs,  we can conclude that  $\lambda_{c,2}> 2\lambda_{c,1}$. A more detailed theoretical exploration of these morphological transitions presents an interesting opportunity for future work.

\section{Discussion\label{sec6:discussions}}

In many practical contexts it is desirable to predict growth conditions which avoid the formation of chimneys in growing mushy layers. In cells that are sufficiently wide to permit the formation of multiple chimneys,  experimental and theoretical models have been applied to infer a critical value R$_m^c$ beyond which the mushy-layer mode convection will occur~\cite[e.g.][]{Fowler:1985, Worster:1992b, ChenLuYang:1994}. However, the cell width $\lambda$ provides an additional constraint on the existence of convective instability. In particular, \cite{Wellsetal:2010} studied the evolution of finite amplitude states of convection with chimneys and showed that as $\lambda$ is decreased the chimneys are extinguished via a saddle-node bifurcation at some critical $\lambda_c(\mathrm{R_m})$. The present work provides the first experimental observation of this finite-sample-size effect on buoyancy-driven convection in the solidification of mushy layers. 

We observe a discernible transition of the flow field in the liquid region overlying the mushy layer as mush-Rayleigh number R$_m$ and cell width are varied. The flow instability is systematically quantified by the sample spacing $\lambda$ and R$_m$ for a variety of liquid concentrations. In Fig. \ref{fig5} we showed two sets of data from experiments conducted at two different cooling rates. At a high cooling rate, the resulting value of R$_m$ and the chimney spacing  $\lambda$ (indicated by the upward triangles in figure \ref{fig4}b)  characterize a stable, no-chimney state (Fig. \ref{fig5}a). A small reduction of the cooling rate increases the value of R$_m$ (shown by the downward triangles in figure \ref{fig4}b), and causes a flow instability in the mushy layer  that results in a new state dominated by the formation of vigorous chimney plumes in the liquid region (Fig. \ref{fig5}b). Using the same nominal experimental control parameters,  the critical Rayleigh number in different runs varies less than $30\%$, as is also evident from the data shown in Fig. \ref{fig6}. Thus, these experimental studies provide reliable results and a general framework for direct comparisons with the prior numerical and theoretical studies, revealing that the system becomes less stable to the onset of flow if either the cell width, Rayleigh number or liquid concentration are increased.

In determining the unknown reference permeability $\Pi_0$ of the mushy layer, we compared experimental and  numerical results for the instability of the mushy-layer mode of convection summarized in Fig. \ref{fig4}. Good comparison can be achieved using a single value of the permeability $\Pi_0=1.6{\times}10^{-5}$ cm$^2$ across the entire experimental range of $C_0$. This suggests that the reference permeability is relatively insensitive to changes in the liquid concentration. It is interesting to compare our estimate of the reference permeability  in our quasi-two dimensional experiment to previous estimates.  For example, for solidification in a fully three-dimensional experimental cell \citet{TaitJaupart:1992}  estimated a mean permeability value  of $\Pi=8\times 10^{-4} \,\mathrm{cm}^2$ based on the Carmen-Kozeny relation and solid fractions inferred from measured temperature and salinity, whilst \cite{NeufeldWettlaufer:2008b} estimated a reference permeability $\Pi_0=1.75\times 10^{-3} \,\mathrm{cm}^2$ by comparing to a linear stability threshold for the onset of shear enhanced convection.
Finally, \cite{ChenChen:1991} estimated a significantly lower value of the mean permeability $\Pi=2.4\times 10^{-5} \,\mathrm{cm}^2$ based on the Carmen-Kozeny relation and solid fractions measured via computed tomography. However, this corresponds to a permeability both after the onset of convection and measured two days after the end of the experiment, allowing the system to evolve away from the initial value of permeability before convective onset. Hence, the difference in magnitude of the inferred reference permeability between our experiments and those of  \cite{TaitJaupart:1992} and  \cite{NeufeldWettlaufer:2008b} suggest that the permeability is significantly lower for solidification in a quasi two-dimensional cell than for a three-dimensional cell.

Through this dynamical constraint of the permeability, the theoretical model for finite-amplitude convecting states successfully captures the form of the observed stability boundary to within the inter-experimental variability of one increment in sample width. This suggests that the theoretical framework of~\cite{Wellsetal:2010} may provide a useful tool in future studies to investigate how chimney formation is influenced by additional factors, such as coupling to flow in the overlying liquid, in a fully nonlinear setting.


\bibliographystyle{jfm}

\end{document}